# Performance of the Mechanical Structure of the SST-2M GCT Proposed for the Cherenkov Telescope Array

**Jean-Laurent Dournaux [1], Jean-Michel Huet, Delphine Dumas, Jean-Philippe Amans, Gilles Fasola, Philippe Laporte, Jean-Jacques Bousquet**
*GEPI. Observatoire de Paris, CNRS, Université Paris Diderot*
*5, place Jules Janssen - 92190 Meudon, France*
*E-mail:* `jean-laurent.dournaux@obspm.fr, delphine.dumas@obspm.fr`

**Hélène Sol**
*LUTH. Observatoire de Paris, CNRS, Université Paris Diderot*
*5, place Jules Janssen - 92190 Meudon, France*
*E-mail:* `helene.sol@obspm.fr`

**For the CTA Consortium[2]**
*www.cta-observatory.org*

The Cherenkov Telescope Array (CTA) consortium aims to create the next generation Very High Energy gamma-ray observatory. It will be devoted to the observation of gamma rays over a wide band of energy, from 20 GeV to 300 TeV. Three different classes, Large, Medium and Small Size Telescopes, are foreseen to cover the low, intermediate and high energy regions, respectively. The energy range of the Small Size Telescopes (SSTs) extends from 1 TeV to 300 TeV. Among them, the Gamma-ray Cherenkov Telescope (GCT), a telescope based on a Schwarzschild-Couder dual-mirror optical design, is one of the prototypes under construction proposed for the SST sub-array of CTA. This contribution focuses on the mechanical structure of GCT. It reports on last progress on the mechanical design and discusses this in the context of CTA specifications. Recent advances in the assembly and installation of the opto-mechanical prototype of GCT on the French site of the Paris Observatory are also described.

---

[1]Speaker

[2] Full consortium author list available at www.cta.observatory.org

1.　Introduction

The Cherenkov Telescope Array (CTA) consortium aims to create the next generation Very High Energy Telescope Array devoted to the observation of gamma rays, from 20 GeV to 300 TeV. Two sites, in the northern hemisphere and in the southern hemisphere, are foreseen, allowing a viewing of the whole sky. Given the wide energy range to be covered, three different classes of telescopes associated to three sizes and energy ranges are therefore proposed. Among them, the Small Size class of Telescopes (SSTs) are devoted to the highest energy region from a few Tev to 300 TeV. About 70 of these telescopes are foreseen in the southern array.

In order to propose a SST structure for the CTA array, the Observatoire de Paris started in 2011 design studies of SST-GATE, a four-meter prototype telescope based on the dual-mirror Schwarzschild-Couder (SC) optical design from preliminary studies developed by the University of Durham [1, 2]. In 2014, the UK-US-NL-German consortium in charge of the development of the camera CHEC (Compact High Energy Camera) [3] joined the SST-GATE team in order to propose GCT (Gamma-ray Cherenkov Telescope), an end-to-end telescope. In spring 2015, the mechanical structure of GCT was assembled on the Meudon site of the Observatoire de Paris.

The SC optical design has never been implemented before in gamma-ray astronomy and is an innovative and interesting alternative to the one-mirror Davis-Cotton design commonly used in ground-based Cherenkov astronomy. This design is particularly advantageous for the SSTs, which require a wide FoV, about 10° [4, 5]. It also allows a compact structure, lightweight camera and enables good angular resolution across the entire FoV while allowing a reduction of the focal length and hence of the physical pixel and overall camera size.

This paper reports on the latest progress on the mechanical structure of GCT as well as on performance of the telescope regarding the CTA specifications.

2.　Telescope structure

**2.1　Overview**

GCT has been designed with a desire to provide a mechanical structure as simple and as light as possible, to ease the mounting and maintenance phase and to decrease manufacturing costs by using commercial-of-the-shelf (COTS) modules and similar systems in the telescope. The mechanical structure of the telescope is detailed in [5] and [6]. The telescope is made up of the four main systems outlined below:
- The tower. The tower is a COTS tube equipped with two flanges. It provides a mechanical interface between the telescope and the foundation and also supports the mass of the telescope.
- The alt-azimuth system (AAS). GCT is equipped with an alt-azimuth mount allowing it to move in azimuth in the range ± 270° and in elevation from 0 to 90°. The AAS is a simple and modular structure, which consists in the fork, the drive systems and the bosshead. It has been designed in order to be similar for both azimuth and elevation subsystems [6].
- The optical support structure (OSS) and the counterweight. The OSS consists in the mast (COTS tubes), the dish M1, the top dish (or dish M2), the bottom dish and the camera removal system. Using two distinct subsystems for supporting the M1 panels

- (dish M1) and for supporting the mast (bottom dish) improves the deflection off-axis of the M1 and allows to use a lighter structure for the dish M1 whereas the bottom dish, heavier, does not have to handle the symmetry of the mirror any more.
- The optical elements. The primary mirror is tessellated and consists of 6 identical panels whereas the secondary mirror is monolithic [7]. The camera is developed by the CHEC consortium [3].

The complete and assembled mechanical structure is shown in Figure 1. The majority of the structural frame is made of standard carbon steel E355. Only the top dish is made of aluminium grade 6. The mechanical parts of the telescope are painted in red (RAL 3016) which offers the best compromise between the reduction of background light during gamma-ray observation and the solar radiation heating. An additional device is planned to clamp the telescope in its stow position, at 0°, on the top dish. Final design of the telescope counts for a total mass of about 7.8 tons which is lighter than other SSTs [5].

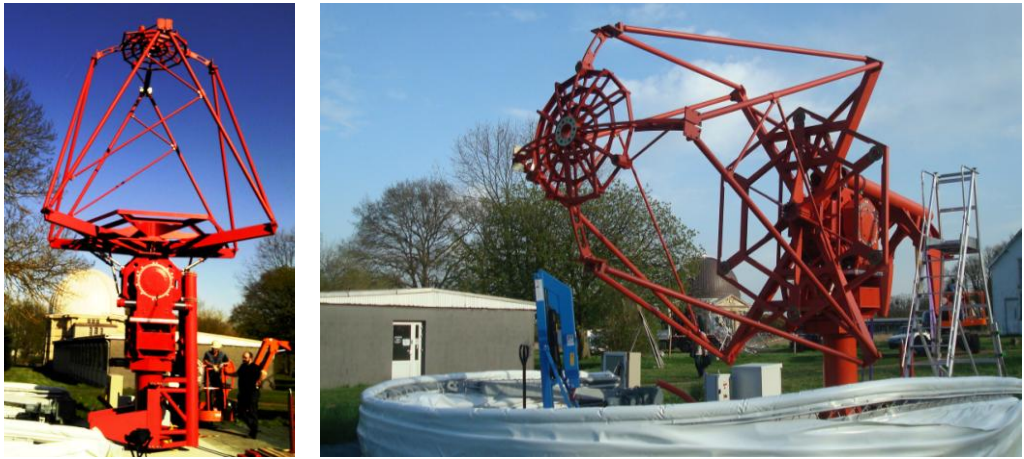

*Figure 1: Mechanical structure of GCT (April 2015) with different elevation angles.*

## 2.2 Manufacturing and assembly

The manufacturing of the different subsystems constituting the mechanical structure was subcontracted to industries between 2012 and 2015. Manufacturing times per subsystem are from two (tower, counterweight), three (OSS) to four (AAS) months. The assembly of the mechanical structure was held on Meudon's site of the Observatoire de Paris in spring 2015. The pre-assembly of the AAS with a perpendicular setting system required two full-time equivalents (FTE) for two weeks and was realized in an integration room. The assembly of the mechanical structure from the pre-assembled subsystems required four FTE for two days and was directly made on site. The only specific materials required for this last purpose were a truck crane, a cherry picker and a forklift.

Real and complete cost for the mechanical structure of the prototype leads to 245 k€ (VAT free). This cost would reduce to 158 k€ for the mass production.

## 3. Performance analysis

Environmental conditions within which CTA must operate are broadly grouped as observing conditions and survival conditions. In observing conditions, the SST must be able to observe with a pointing accuracy of 1 mrad for elevation angles between 20° and 60° and can be subjected to gravity loads, thermal gradients below 17.5°C and below-moderate wind speeds (< 36 km/h). This accuracy is derived into optics displacements expressed themselves in deviation (component parallel to the optical axis), decentring (component perpendicular to the optical axis) and tilt [8]. **Erreur ! Source du renvoi introuvable.**In survival conditions, the telescope is in a stow position and shall not be damaged. It can be subject to wind speeds up to 150 km/h with gusts at 200 km/h as well as snow and ice loads as specified in [9]. A finite element (FE) model, detailed in [6], has been developed to determine if the design fulfils these requirements. FE model's results are expressed in the local coordinate system related to the OSS in which x is along the elevation axis, z is along the optical axis and y is deduced by a right-hand rule.

Table 1 summarizes the first three oscillation modes computed for the stow position and for several elevation angles. When the eigenmode does not exist at a given elevation angle or at a given configuration, it is noted as being non-existent. These data demonstrate that the first eigenfrequency occurs at a frequency similar to other CTA dual-mirror telescope [10]. These modes are plotted in Figure 2.

| Mode | Description | Stow | 0° | 20° | 60° | 90° |
|---|---|---|---|---|---|---|
| 1 | Rotation of the OSS around azimuth axis | 3.7 | 3.4 | 3.4 | n.e. | n.e. |
| 2 | Bending of the telescope around y | n.e. | 4.0 | 3.9 | 3.6 | 3.6 |
| 3 | Bending of the telescope around x | 8.7 | 4.9 | 4.7 | 4.5 | 4.4 |

*Table 1: Summary of the first three eigenmodes of GCT. Frequencies are given in Hz (n.e: non-existent).*

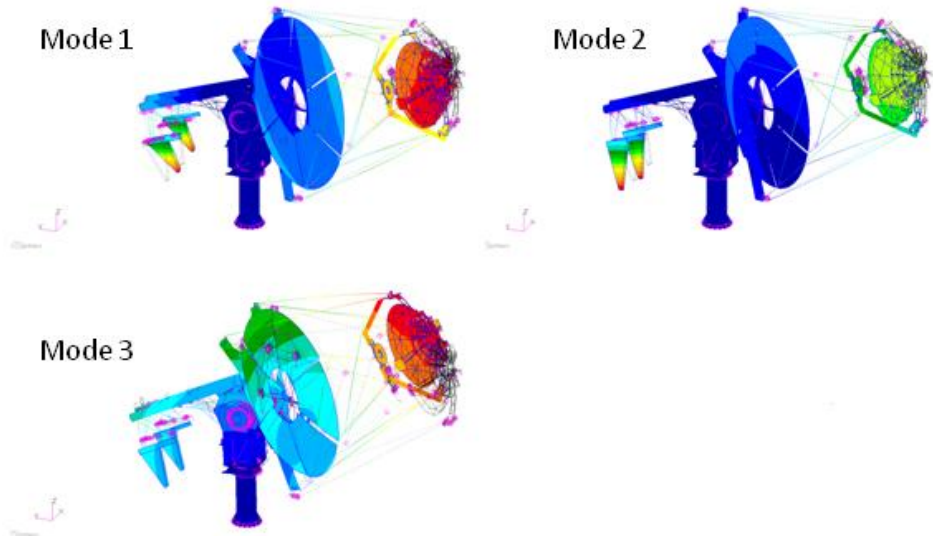

*Figure 2: First three eigenmodes computed for 20° elevation.*

Computed displacements and rotations of the centres of gravity of the optical elements in observing conditions are compared to the allocations deduced from CTA specifications for the structural (gravity with or without wind) and the thermomechanical case in Table 2 and Table 3 respectively. These data demonstrate that the design largely fulfils specifications in observing conditions. The deformed shape of the telescope for 20° elevation and a front wind is plotted in Figure 3.

When the telescope is in survival conditions, stresses remain below the yield stresses of the constitutive materials scaled by a security factor, showing that no damage would occur. Maximal stresses are located in the top dish and in the counterweight.

|  | Dec M1M2 | Dec M2Cam | Dev M1M2 | Dev M2Cam | Tilt x M1M2 | Tilt y M1M2 | Tilt x M2Cam | Tilt y M2Cam |
|---|---|---|---|---|---|---|---|---|
| Spec (PtV) | 8.5 | 8.5 | 7.75 | 7.75 | 900 | 900 | 900 | 900 |
| Gravity | 4.3 | 0.1 | 0.9 | -0.7 | 26 | ≤ 1 | -3 | -18 |
| Grav. + front wind | 4.2 | 0.1 | 1.0 | -0.7 | 26 | ≤ 1 | -3 | -14 |
| Grav. + back wind | 4.3 | 0.1 | 0.9 | -0.7 | 29 | 14 | -1 | -48 |

*Table 2: Maximal gravity-induced and wind-induced on-axis deformations of optical components centres of gravity in observing conditions. Displacements are given in mm whereas rotations are given in arcsec.*

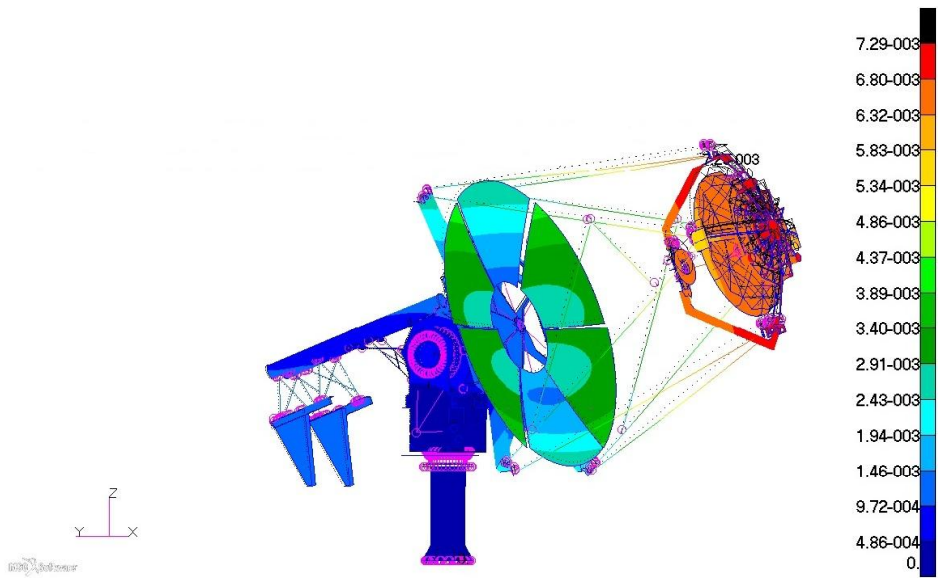

*Figure 3: Plot of the displacements computed for 20° elevation and a 36 km/h front wind. The non-deformed shape appears in dotted lines. The displacements are computed in magnitude and are plotted in meters.*

|  | Dec M1M2 | Dec M2Cam | Dev M1M2 | Dev M2Cam | Tilt x M1M2 | Tilt y M1M2 | Tilt x M2Cam | Tilt y M2Cam |
|---|---|---|---|---|---|---|---|---|
| Spec (PtV) | 1 | 1 | 1.1 | 1.1 | 47 | 47 | 47 | 47 |
| Thermomechanical | 0.02 | 0.1 | 0.9 | -0.5 | 9.5 | -5.4 | -19.9 | 12.8 |

*Table 3: Maximal temperature-induced on-axis deformations of optical components centres of gravity in observing conditions. Displacements are given in mm whereas rotations are given in arcsec.*

4. Conclusion

GCT is an end-to-end telescope, proposed as a SST for the future Cherenkov Telescope Array (CTA). It provides a light, low-cost and simple mechanical structure, which has been successfully assembled in spring 2015. Procurement of the different subsystems takes from two to four months, which allows fast construction compatible with CTA construction timetable. The alt-azimuth subsystem requires a pre-assembly which can be made by two FTE in two weeks and the total structure can be assembled by four FTE in two days. Performance analysis of the telescope, made by FEA, shows that the prototype is planned to observe with the specified accuracy in the observing environment conditions defined by the CTA Consortium and that no damage would occur to the structure under the survival conditions.

## Acknowledgements

The authors gratefully acknowledge the council of the Region Ile-de-France, the CNRS and the Observatoire de Paris for supporting the SST-GATE project; the suppliers ACO, Ambos and Alsyom for their quality collaboration in the manufacturing of the telescope. We also gratefully acknowledge support from the agencies and organizations listed under Funding Agencies at this website: http://www.cta-observatory.org/.